%% file: main.tex
\documentclass[%
 aip,
 amsmath,amssymb,
reprint
]{revtex4-2}

\usepackage{graphicx,color}% Include figure files
\usepackage{dcolumn}% Align table columns on decimal point
\usepackage{bm}% bold math
\usepackage[utf8]{inputenc}
\usepackage[T1]{fontenc}
\usepackage{mathptmx}
\usepackage{etoolbox}
\usepackage{CJK}
\usepackage{makecell}
\def\thickhline{\noalign{\hrule height1.2pt}}
\usepackage{verbatim}
\usepackage{anyfontsize}
\makeatletter
\def\@email#1#2{%
 \endgroup
 \patchcmd{\titleblock@produce}
  {\frontmatter@RRAPformat}
  {\frontmatter@RRAPformat{\produce@RRAP{#1\href{mailto:#2}{#2}}}\frontmatter@RRAPformat}
  {}{}
}%
\makeatother
\begin{document}

\preprint{AIP/123-QED}

\title[A Steady Loop Current Does Not Radiate]{A Steady Loop Current Does Not Radiate}
% Force line breaks with \\
%\author{
%Shengchao Alfred Li (\begin{CJK*}{UTF8}{gbsn}李盛超%\end{CJK*})$^1$, %\and 
%Charlotte Jingyang Li (\begin{CJK*}{UTF8}{gbsn}李景洋%\end{CJK*})}
%\email[$^1$ Corresponding author; Email: ]%{Shengchao.Li@gmail.com.}

\author{
Shengchao Alfred Li (\begin{CJK*}{UTF8}{gbsn}李盛超\end{CJK*})}, 
\thanks{Corresponding author: \texttt{Shengchao.Li@gmail.com}}
\affiliation{An amateur scientist in Potomac, Maryland, USA.}
\author{Charlotte Jingyang Li (\begin{CJK*}{UTF8}{gbsn}李景洋\end{CJK*})}
\affiliation{An amateur scientist in Potomac, Maryland, USA.}
\affiliation{Current address: University of Maryland, College Park, Maryland, USA. (First-year undergraduate)}
%\email[Corresponding author; Email: ]{Shengchao.Li@gmail.com.}
\date{Aug. 08, 2023; Latest updated: Oct. 08, 2025}

%\email[$^2$]{Current affiliation: University of Maryland}

%\affiliation{Independent researcher, Potomac, Maryland, USA.}
%\altaffiliation[Also at ]{Potomac, Maryland, USA.}%Lines break automatically or can be forced with \\
%\author{Charlotte Jingyang Li (\begin{CJK*}{UTF8}{gbsn}李景洋\end{CJK*})}
%\affiliation{Independent researchers, Potomac, Maryland, USA.}
%\altaffiliation[Also at ]{University of Maryland}
%\\This line break forced with \textbackslash\textbackslash
%}%


\begin{abstract}
According to classical electrodynamics, a steady loop current does not radiate. Edward, Kenyon, and Lemon show that the present-time approximation of the retarded electric field (the approximation of the magnetic one is trivial) of a steady loop current can be partitioned into a loop integral of a static field expression and a loop integral of an exact (also called total, full, or perfect) differential. Because the latter is zero, no radiation is emitted. Inspired by their work, we do the same for the retarded electric and magnetic fields without approximation, and show that for a steady loop current, the loop integrals of the static field expressions with and without approximation are exactly equal, recovering Coulomb's law and the Biot-Savart law for a steady loop current.
\end{abstract}
\maketitle

\section{Introduction}\label{sec:introduction}
It is well known that a steady loop current does not radiate because the scalar and vector potentials are constant \cite{griffiths1999introduction} (Ref. [\onlinecite[p.444]{griffiths1999introduction}]). Some authors have also illustrated this result from the moving point charge view \cite{panofsky1962classical, zapolsky1988electric, jackson1998classical, mcdonald2001whyDA} (Ref. [\onlinecite[p.370]{panofsky1962classical}], Ref. [\onlinecite[Sect. V]{zapolsky1988electric}],  Ref. [\onlinecite[problem 14.24]{jackson1998classical}]), which involves the retardation-based Li\'enard-Wiechert potentials and fields.

However, these authors assume a circular loop current with same-speed point charges in the current \cite{panofsky1962classical, zapolsky1988electric, jackson1998classical, mcdonald2001whyDA}.  So far the literature still misses the analysis of the retarded fields of a general loop current (arbitrarily shaped and with arbitrary, but steady charge speed) from the moving point charge view (see TABLE \ref{tab:missingness}). 

It is known that working with retardation (involving retarded quantities) is not straightforward \cite{melia2001electrodynamics} (Ref. [\onlinecite{melia2001electrodynamics}, p.100]). Historically, authors have used Taylor expansion to come up with the present-time approximations\cite{o1938electromagnetic, o1965electromagnetic, landau1975classical,jackson1998classical} ( Ref. [\onlinecite{o1938electromagnetic, o1965electromagnetic}, Ritz's work], Ref. [\onlinecite{landau1975classical,jackson1998classical}, Darwin's Lagrangian]) of the Li\'enard-Wiechert fields  up to $1/c^2$, where $c$ is the speed of light in vacuum. Approximations up to $1/c^5$ have also been worked out\cite{page1918moving, page1922introduction}. 

\begin{table}\label{tab:missingness}
\caption{Current literature about why a loop current does not radiate from the moving point charge view.}
%\begin{ruledtabular}
\begin{tabular}{c|c|c}
%\rowcolor{lightgray!40}
&\multicolumn{2}{c}{from the moving point charge view}\\
\cline{2-3}
%\rowcolor{lightgray!40}
\makecell{loop current's shape\\and charge speed}& \makecell{with present-time,\\but approximate\\field expressions} & \makecell{with exact, \\but retarded\\field expressions}\\
\thickhline
\makecell{circular,\\same-speed} & \makecell{not very\\interesting} & \makecell{Ref. [\onlinecite{panofsky1962classical, zapolsky1988electric, jackson1998classical, mcdonald2001whyDA}]\\(with point charges)}\\
\hline
\makecell{general shape,\\arbitrary-speed} & \makecell{Ref. [\onlinecite{edwards1976continuing}]\\(with fluid model)}& \makecell{this paper\\(with fluid model)}\\
\end{tabular}
%\end{ruledtabular}
\end{table}

\begin{figure}
    \centering
    \scalebox{0.7}{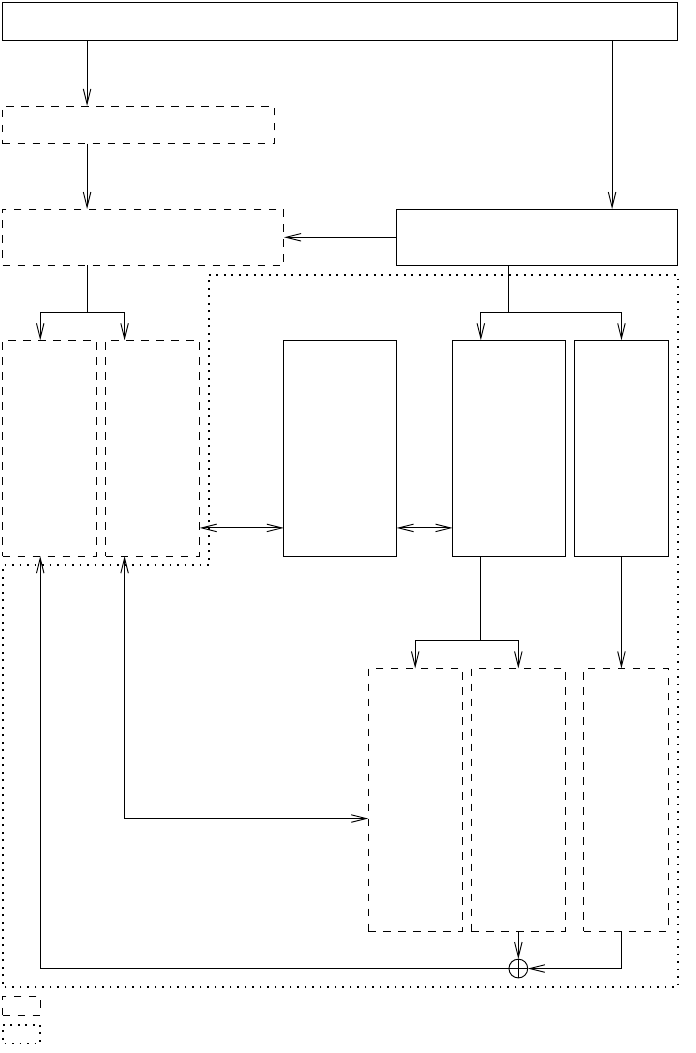}
    \caption{This paper in the context of current literature.}
    \label{fig:compare}
\end{figure}

With the present-time approximations up to $1/c^2$ and for a general loop current, Edward, Kenyon, and Lemon\cite{edwards1976continuing} show that the electric field of a moving point charge is composed of a static field and an exact differential (obtainable directly from Darwin's approximate potentials in the Coulomb gauge, as pointed out by McDonald \cite{mcdonald2019electric}),  thus the integral of the latter over the loop current is zero. Their work uses the fluid model \cite{westgard1995electrodynamics}, also known as the single-velocity continuum model \cite{bobroff1959independent}. With this model, a moving, continuous linear charge distribution is assumed for the loop current, while for a small charge element of the current, the retarded field expressions for a moving point charge are used. Their experiment showing a deviation from the theory has been challenged \cite{shishkin2002investigation} and is irrelevant here.

In this paper, we fill the void (see TABLE \ref{tab:missingness}) and show that with the fluid model, for a general loop current, the retarded electric field can also be partitioned into a loop integral of a static field expression and a loop integral of an exact differential, and thus no radiation is emitted. In addition, we show that it is also the case for the retarded magnetic field. 

In FIG. \ref{fig:compare}, we depict the relationship of this paper to current literature, especially to Edward, Kenyon, and Lemon's work\cite{edwards1976continuing}.

We hope our work will help students to have a deeper understanding of the retarded fields of an arbitrarily moving point charge. %It may also help people have a deeper understanding of radiation in electron beams \cite{??}.

\section{Integration elements of the steady loop current}

For the purpose described in Sect. \ref{sec:introduction}, with the fluid model, for a general loop with a steady current, we need to write out the retarded electric and magnetic fields as loop integrals of the fields of the moving charge elements. Thus, we need to find the infinitesimal integration elements in terms of the steady current, shown in Eq. (\ref{dq}) at the end of this section.

\begin{figure}
    \centering
    \scalebox{0.7}{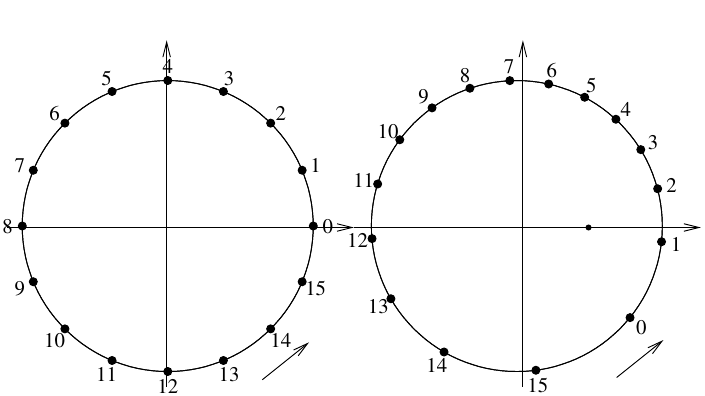}
    \caption{(a) A circular steady loop current with labeled point charges that are evenly spaced. (b) The retarded positions of the labeled point charges seen by the observer at position $obs$. Plots are generated with the computer code listed in Section \ref{sec:code}. Note that if the moving charges are negative, $I$ takes negative value.}
    \label{fig:byN}
\end{figure}

Refer to FIG. \ref{fig:byN}(a). To start with something simple, we study the elements of a steady current in an electrically neutral, circular, thin wire loop, with all moving charges having the same speed. With ``neutral'' we mean that there is an equal number of positive and negative charges. This ideal thin circular current may be modeled from realistic steady superconductor currents\cite{file1963observation, shishkin2002investigation}. Let the radius of the circle be $a$ and the total charge of the current be $q$, with the understanding that $q$ is negative if the current consists of electrons. Let the charge speed be $v$ and $T=2\pi a/v$ be the time interval for a point charge to make a round trip. We deduce that the current $I=q/T$. Note that if the moving charges are negative, $I$ is negative. Because a electric charge is invariant regardless of its motion\cite{schwartz1972principles} (Ref. [\onlinecite{schwartz1972principles}, p.121, last paragraph]) and the loop is neutral, we deduce that the total static charge in the loop is $-q$.

As depicted in FIG. \ref{fig:byN}(a), at present time $t$, we selectively label evenly spaced $N$ point charges in the current. We label them with numbers $0, 1, \cdots, N-1, N$, where $N$ and $0$ label the same point charge. In normal situations, such as with a metal wire ring, this is feasible because there are an enormous number of point charges in the current. Due to symmetry, we know that the amount of moving charge between two neighboring labels is $\delta q=q/N$.  It is straightforward to deduce that, at time $t+\delta t\stackrel{\text{def}}{=}t+T/N$, each labeled point charge arrives at the position the next charge happened to be at time $t$. Thus, we have the relationship
\begin{align}
    \delta q=I\delta t.\label{deltaqdeltat}
\end{align}

Let an observer sit at an arbitrary position ${\mathbf r}_{obs}$, labeled $obs$ in FIG. \ref{fig:byN}(b).  Because fields propagate with light speed $c$, for the $k$'th labeled point charge, what the observer sees is not its current position but its retarded position ${\mathbf r}_k^\prime(t_k^\prime)$ at the retarded time 
\begin{align}
t_k^\prime&=t-|{\mathbf r}_{obs}-{\mathbf r}_k^\prime(t_k^\prime)|/c.\label{retarded}
\end{align}

\emph{In this paper, we use the prime symbol ``$^\prime$'' to indicate quantities related to the source charge, which, unless explicitly stated, are retarded quantities at the retarded time due to the finite propagation speed of light from the source charge to the observer.} This convention is consistent to popular textbooks \cite{schwartz1972principles, jackson1998classical,griffiths1999introduction}.

In FIG. \ref{fig:byN}(b) we show the retarded positions of the $N$ labeled point charges the observer sees at time $t$, calculated for a set of parameters with the computer code listed in Section \ref{sec:code}. An important fact is that the charge the observer sees at time $t$ between two neighboring labeled point charges ${\mathbf r}_{k-1}^\prime(t_{k-1}^\prime)$ and ${\mathbf r}_{k}^\prime(t_{k}^\prime)$, denoted by $\delta q_k^\prime$, is just the same charge between ${\mathbf r}_{k-1}(t)$ and ${\mathbf r}_{k}(t)$. Thus, we have
\begin{align}
\delta q_k^\prime=\delta q,\label{deltaqprime}
\end{align}
because of the invariance property of a single moving charge discussed above. 

We now transfer the labels on the point charges at positions ${\mathbf r}_k^\prime(t_k^\prime)$, $k=0, 1, \cdots, N$ to the loop itself and fix them in place, and study how the charges move relative to these fixed positions. Because the labels are now fixed, we can drop the time and write the positions as ${\mathbf r}_k^\prime$, $k=0, 1, \cdots, N$, with ${\mathbf r}_0^\prime={\mathbf r}_N^\prime$. These positions divide the circle into N arcs, which in general are not equal in length (FIG. \ref{fig:byN}(b)). At time $t+\delta t=t+T/N$, the point charge previously seen by the observer at ${\mathbf r}_{k-1}^\prime$ is now seen at ${\mathbf r}_{k}^\prime$, whose retarded time at these two places are $t-|{\mathbf r}_{obs}-{\mathbf r}_{k-1}^\prime|/c$ and $t+\delta t-|{\mathbf r}_{obs}-{\mathbf r}_{k}^\prime|/c$, respectively. This is true for all $k=1, 2, \cdots, N$. If we sum up the $N$ time-interval $\delta t$'s for the $N$ arcs, we get the period $T$.

\begin{figure}
    \centering
    \scalebox{0.7}{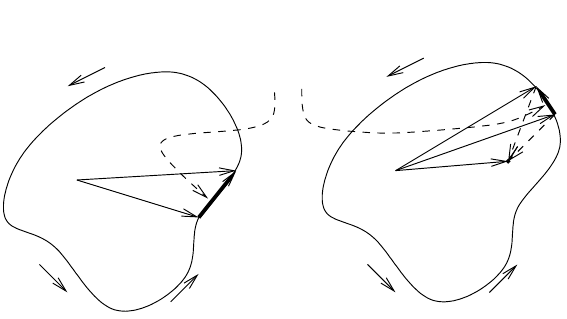}
    \caption{A general loop carrying steady current $I$. Note that if the moving charges are negative, $I$ takes negative value. The loop is not necessarily in a plane. It can be any 3-dimensional closed loop. $O$ is the origin of the coordinate system and $obs$ is the position of the observer. (a) A present-time segment $\delta {\mathbf l}_k$ of the loop is labeled. (b) The corresponding retarded segment $\delta{\mathbf l}_k^\prime$ of the same loop, containing the same charge, is labeled. The offset between $\delta{\mathbf l}_k^\prime$ and $\delta {\mathbf l}_k$ are greatly exaggerated for normal situations. They are labeled on the same loop, not on two separate loops.}
    \label{fig:model}
\end{figure}

When $N$ is large, the arcs can be approximated with segments. Our analysis so far is general enough to apply to steady loop currents with arbitrary shapes and arbitrary, but steady, charge speeds at different positions of the loop. We depict in FIG. \ref{fig:model}(b) a general loop current, where the $k$'th segment is denoted by $\delta {\mathbf l}_k^\prime={\mathbf r}_{k}^\prime-{\mathbf r}_{k-1}^\prime$, $k=1$, $2$, $\cdots$, $N$. Since a point charge at the position ${\mathbf r}_{k-1}^\prime$ at time $t-|{\mathbf r}_{obs}-{\mathbf r}_{k-1}^\prime|/c$ moves to ${\mathbf r}_{k}^\prime$ at time $t+\delta t-|{\mathbf r}_{obs}-{\mathbf r}_{k}^\prime|/c$, we have
\begin{align}
\delta {\mathbf l}_k^\prime&={\mathbf r}_{k}^\prime-{\mathbf r}_{k-1}^\prime\stackrel{\text{def}}{=}{\mathbf v}^\prime\delta t_{k}^\prime, \label{deltal}
\end{align}
where 
\begin{align}
\delta t_k^\prime=&(t+\delta t-|{\mathbf r}_{obs}-{\mathbf r}_{k}^\prime|/c)-(t-|{\mathbf r}_{obs}-{\mathbf r}_{k-1}^\prime|/c)\notag\\
=&\delta t-(|{\mathbf r}_{obs}-{\mathbf r}_{k}^\prime|-|{\mathbf r}_{obs}-{\mathbf r}_{k-1}^\prime|)/c,\label{deltat}
\end{align}
and ${\mathbf v}^\prime$ is defined as the average velocity of the charge traveling along $\delta {\mathbf l}_k^\prime$. 
%This equation is as simple as it is literally, showing that the segment the charge has traveled during a time interval is the product of the time interval and the charge's average velocity.

In the following, we use the fluid model \cite{bobroff1959independent}. If we let $N\rightarrow \infty$ and drop the subscript $_k$, $\delta q$, $\delta q_k^\prime$, $\delta t$, $\delta t_k^\prime$, $\delta {\mathbf l}_k^\prime$, and ${\mathbf r}_k^\prime$ become $dq$, $dq^\prime$, $dt$, $dt^\prime$, $d{\mathbf l}^\prime$ and ${\mathbf r}^\prime$, and Eqs. (\ref{deltaqprime}) and (\ref{deltaqdeltat}) become
\begin{align}
dq^\prime=dq=Idt,\label{dqdt}
\end{align}
while Eq. (\ref{deltal}) becomes
\begin{align}
d{\mathbf l}^\prime&={\mathbf v}^\prime dt^\prime, \label{dl}
\end{align}
consistent with the definition of ${\mathbf v}^\prime$ in the textbooks\cite{panofsky1962classical, podolsky1969fundamentals, schwartz1972principles, heald1995classical} (Ref. [\onlinecite{panofsky1962classical}, Eq. (20-1)], [\onlinecite{schwartz1972principles}, the Eq. above Eq. (6-3-2)], [\onlinecite{podolsky1969fundamentals}, Eq. (19.24)], [\onlinecite{heald1995classical}, Eq. (8.39)]). For a steady current, the present-time velocity of charges \emph{in the same segment} $d{\mathbf l}^\prime$ is the same as the retarded velocity ${\mathbf v}^\prime$ found there. However, we reserve the symbol ${\mathbf v}$ for the present-time velocity \emph{of the same charge}, which, in general, is not in the same segment $d{\mathbf l}^\prime$ and has ${\mathbf v}\neq{\mathbf v}^\prime$.

In addition, $|{\mathbf r}_{obs}-{\mathbf r}_{k}^\prime|-|{\mathbf r}_{obs}-{\mathbf r}_{k-1}^\prime|$ becomes $-{\mathbf n}^\prime\cdot d{\mathbf l}^\prime$, where ${\mathbf n}^\prime=({\mathbf r}_{obs}-{\mathbf r}^\prime)/|{\mathbf r}_{obs}-{\mathbf r}^\prime|$ is the direction of ${\mathbf r}_{obs}-{\mathbf r}^\prime$. With the help of this equation and Eq. (\ref{dl}), Eq. (\ref{deltat}) becomes
\begin{align}
dt^\prime&=dt+{\mathbf n}^\prime\cdot d{\mathbf l}^\prime/c\notag\\
&=dt+dt^\prime({\mathbf n}^\prime\cdot {\mathbf v}^\prime)/c,\label{dt}
\end{align}
thus\cite{schwartz1972principles, melia2001electrodynamics}
\begin{align}
    dt^\prime=\frac{dt}{1-{\mathbf n}^\prime\cdot{\mathbf v}^\prime/c}=D^\prime dt,\label{dtprime}
\end{align}
where
\begin{align}
    D^\prime=&\frac{1}{1-{\mathbf n}^\prime\cdot{{\mathbf v}^\prime}/{c}},
\end{align}
which, according to Podolsky and Kunz, is the Doppler factor\cite{podolsky1969fundamentals}. For this quantity, Davidson invokes acoustics \cite{davidson2019introduction} ,  Griffiths says ``reminiscent of the Doppler effect'' \cite{griffiths1999introduction}, and Zangwill uses the name ``Doppler-like factor'' \cite{zangwill2013modern}. It has many other names, including ``volumetric correction'' \cite{davidson2019introduction}, ``direction factor'' \cite{westgard1995electrodynamics}, and for its inverse, Jacobian \cite{melia2001electrodynamics} (of a space transformation), or ``shrinkage factor'' \cite{melia2001electrodynamics}. 

Using Eqs. (\ref{dqdt}), (\ref{dl}), and (\ref{dtprime}), our analysis finally shows that for a general steady loop current with present-time current $I$, which is the same for the whole loop, the retarded charge in the source segment (or source element) $d{\mathbf l}^\prime$ is
\begin{align}
dq^\prime&=dq=Idt=I\frac{dt^\prime}{D^\prime}=I\frac{d{\mathbf l}^\prime}{D^\prime {\mathbf v}^\prime}\notag\\
&=I\frac{dl^\prime}{D^\prime v^\prime},\quad \text{(${\mathbf v}^\prime$ and $d{\mathbf l}^\prime$ share direction)}\label{dq}
\end{align}
where $dl^\prime=|d{\mathbf l}^\prime|$, 
$v^\prime=|{\mathbf v}^\prime|$, $dt^\prime$ is the time interval for a charge to move along the source segment $d{\mathbf l}^\prime$, and ${\mathbf v}^\prime$ is the retarded velocity of the charge that affects the observer at the present time. 

The derivation of Eq. (\ref{dq}) is related to the textbook derivation of the Li\'{e}nard-Wiechert potentials\cite{griffiths1999introduction, panofsky1962classical, jackson1998classical}, but there, the effect (the fields) of a point charge is concentrated by $D^\prime$ due to the Doppler or Doppler-like effect, while here the charge density is diluted by $1/D^\prime$ because the retarded segment is stretched by $D^\prime$ due to retardation of time. 

The retarded current, also subject to this dilution, is defined as 
\begin{align}
I^\prime=dq^\prime/dt^\prime=I/D^\prime.\label{Iprime}
\end{align}
However, it is not very attractive because, unlike $I$, even for a steady current loop, $I^\prime$ varies for different segments of the loop and for observers at different positions.

%These relations tell us that, for a source segment ${d\mathbf l}^\prime$ with charge velocity ${\mathbf v}^\prime$, the effect of retardation, or the effect of finite light speed, is that the retarded charge $dq^\prime$ in it that affects the observer at current time $t$, is scaled by the inverse Doppler factor $1/D^\prime$. If we define retarded current element $I^\prime d{\mathbf l}^\prime$

\section{Fields of a Steady Loop Current without Approximation}\label{sec:Fields}

We set out to find the electric field ${\mathbf E}_L$ and the magnetic field ${\mathbf B}_L$, where the subscript $_L$ stands for ``loop'', of a general steady loop current at an observer position. They are integrals of retarded fields of charge elements of the current. 

To make the expressions concise, without loss of generality, we let ${\mathbf r}_{obs}={\mathbf 0}$. Thus, ${\mathbf r}_{obs}-{\mathbf r}^\prime={\mathbf 0}-{\mathbf r}^\prime=-{\mathbf r}^\prime$ and ${\mathbf n}^\prime=({\mathbf r}_{obs}-{\mathbf r}^\prime)/|{\mathbf r}_{obs}-{\mathbf r}^\prime|=-{\mathbf r}^\prime/r^\prime$, where $r^\prime=|{\mathbf r}^\prime|$.

The retarded fields of a charge element, known as the fields of an arbitrarily moving point charge, or Li\'{e}nard-Wiechert fields according to some authors, are found in almost all electrodynamic textbooks. Many intermediate expressions are also available, from which we choose a pair \cite{jackson1962classical,poission2005electromagnetic, zangwill2013modern, nolting2016theoretical} (Ref.[\onlinecite{jackson1962classical}, Eqs. (14.9)], Ref. [\onlinecite{poission2005electromagnetic}, p.117, p.118], Ref. [\onlinecite{zangwill2013modern}, Eqs. (23.24), (23.25)], Ref. [\onlinecite{nolting2016theoretical}, Eqs. (4.509), (4.511)]) that makes our work easy,
\begin{align}
d{\mathbf E}=&\frac{D^\prime dq^\prime}{4\pi\epsilon_0}\left(\frac{{\mathbf n}^\prime}{r^{\prime 2}}+ \frac{1}{c}\frac{d}{dt^\prime}\left(\left({\mathbf n}^\prime-\frac{{\mathbf v}^\prime}{c}\right)\frac{D^\prime}{r^\prime}\right)\right),\\
d{\mathbf B}=&\frac{D^\prime dq^\prime}{4\pi\epsilon_0c^2}\left(\frac{{\mathbf v}^\prime\times{\mathbf n}^\prime}{r^{\prime 2}} +\frac{d}{dt^\prime}\left(\left(\frac{{\mathbf v}^\prime}{c}\times{\mathbf n}^\prime\right)\frac{D^\prime}{r^\prime}\right)\right).
\end{align}
The second term of each of the two expressions, with the help of Eq. (\ref{dq}), can be written as an exact differential, to give us
\begin{align}
d{\mathbf E}=&\frac{D^\prime }{4\pi\epsilon_0}\frac{{\mathbf n}^\prime}{r^{\prime 2}}dq^\prime+\frac{I}{4\pi\epsilon_0 c}d\left(\left({\mathbf n}^\prime-\frac{{\mathbf v}^\prime}{c}\right)\frac{D^\prime}{r^\prime}\right),\label{dE1}\\
d{\mathbf B}=&\frac{D^\prime }{4\pi\epsilon_0c^2}\frac{{\mathbf v}^\prime\times{\mathbf n}^\prime}{r^{\prime 2}}dq^\prime+\frac{I}{4\pi\epsilon_0c^2}d\left(\left(\frac{{\mathbf v}^\prime}{c}\times{\mathbf n}^\prime\right)\frac{D^\prime}{r^\prime}\right).\label{dB1}
\end{align}

Again, with the help of Eq. (\ref{dq}) and (\ref{Iprime}),
We can express the first terms in a few ways,
\begin{align}
&\frac{D^\prime dq^\prime}{4\pi\epsilon_0}\frac{{\mathbf n}^\prime}{r^{\prime 2}}=\frac{1}{4\pi\epsilon_0}\frac{{\mathbf n}^\prime}{r^{\prime 2}}dQ^\prime=\frac{1}{4\pi\epsilon_0}\frac{{\mathbf n}^\prime}{r^{\prime 2}}\left(I \frac{d{\mathbf l}^\prime}{{\mathbf v}^\prime}\right)\label{retardedCoulomb}\\
&\frac{D^\prime dq^\prime}{4\pi\epsilon_0c^2}\frac{{\mathbf v}^\prime\times{\mathbf n}^\prime}{r^{\prime 2}}=\frac{I^\prime/D^\prime}{4\pi\epsilon_0c^2}\frac{d{\mathbf l}^\prime\times{\mathbf n}^\prime}{r^{\prime 2}}=\frac{1}{4\pi\epsilon_0c^2}\frac{I d{\mathbf l}^\prime\times{\mathbf n}^\prime}{r^{\prime 2}},\label{retardedBfield}
\end{align}
where $dQ^\prime=D^\prime dq^\prime$, the ``effective retarded charge'' \cite{panofsky1951classical} (Ref. \onlinecite[p.197]{panofsky1951classical}), which happens to equal $Id{\mathbf l}^\prime/{\mathbf v}^\prime$, which exposes the way to find the charge in the segment when we apply the ordinary Coulomb's law. Thus, Eqs. (\ref{retardedCoulomb}) and (\ref{retardedBfield}) are also the present-time formulas for a steady current $I$. What has happened is that the retarded charge's density is diluted by the inverse Doppler factor $1/D^\prime$, but its effect is augmented by the Doppler factor $D^\prime$. The two effects cancel each other out.

Interestingly, with similar arguments, we see that for the neutral circular loop current depicted in FIG. \ref{fig:byN}(b), the amount of static charge in $d{\mathbf l}^\prime$, with opposite sign, without being subject to dilution, is $D^\prime$ times more than the retarded moving charge $dq^\prime$. 
Thus, the electric fields of the static charge in the segment $d{\mathbf l}^\prime$, determined by the ordinary (present-time) Coulomb's law, and that of the moving charge in the same segment, determined by Eq. (\ref{retardedCoulomb}), cancel each other out at the observer position. This result agrees with Clausius' assumption that a neutral wire (neutral when there is no current) carrying a steady current is still neutral, that is, it does not exert forces on a static test charge\cite{clausius1877ueber}. This result is also obtained for a long straight wire carrying a steady current through Gauss's law for both thin \cite{zapolsky1988electric} and cylindrical wire\cite{gabuzda1993charge, van2008lorentz} because their present-time linear charge density is still zero (Ref. [\onlinecite{gabuzda1993charge}] provides a mechanism for the surface charge Ref. [\onlinecite{peters1985frame}] asked for; Ref. [\onlinecite{van2008lorentz}] Sect. 2 only \cite{redzic2010comment}). For the latter, although several textbooks mistakenly assume neutral volume charge density (as pointed out by Ref. [\onlinecite{peters1985frame}] [\onlinecite{matzek1968transverse}] [\onlinecite{mcdonald2021charge}]), the non-zero volume charge density induced by self-induced Hall effect (or pinch effect) is compensated by surface charge with the opposite sign\cite{peters1985frame, gabuzda1993charge, van2008lorentz}.

Because loop integrals of the exact differentials are zero, the integration of Eqs. (\ref{dE1}) and (\ref{dB1}) over the loop gives the integration of Eqs. (\ref{retardedCoulomb}) and (\ref{retardedBfield}), or

\begin{align}
    {\mathbf E}_L&=\frac{1}{4\pi\epsilon_0}\oint_L  \frac{{\mathbf n}^\prime}{r^{\prime 2}}\left(I \frac{d{\mathbf l}^\prime}{{\mathbf v}^\prime}\right)\label{E_Lloop},\\
    {\mathbf B}_L&=\frac{1}{4\pi\epsilon_0c^2}\oint_L \frac{I d{\mathbf l}^\prime\times{\mathbf n}^\prime}{r^{\prime 2}}.\label{B_Lloop}
    %=\frac{1}{4\pi\epsilon_0c^2}\oint_L I^\prime D^\prime\frac{d{\mathbf l}^\prime\times{\mathbf n}^\prime}{r^{\prime 2}}
\end{align}
These laws, which might be called the retarded Coulomb's law and the retarded Biot-Savart law for a steady loop current $I$, take the same forms and yield the same results as the ordinary Coulomb's law and the ordinary Biot-Savart law in integration forms. Because the fields fall with the square of the distance $r^\prime$, they are static fields. Thus, a general steady loop current does not radiate.

\section{Previous Results}\label{section:previous}

For comparison purposes, we briefly describe Edward, Kenyon, and Lemon's work here. 
%Note that we set the origin of the coordinates at the observer position, e.g. ${\mathbf r}_{obs}={\mathbf 0}$, where ${\mathbf 0}$ represents the vector with zero length.

%%these two equations are checked to be correct
%The electric and magnetic fields of an arbitrarily moving point charge are \cite{panofsky1962classical, schwartz1972principles},
%\begin{align}
%{\mathbf E}=&\frac{1}{4\pi\epsilon_0}\frac{D^{\prime 3}q}{c^2r^\prime}\left(-{\mathbf a}^\prime+{\mathbf n}^\prime({\mathbf n}^\prime\cdot{\mathbf a}^\prime)+{\mathbf n}^\prime\times\left({\mathbf a}^\prime\times{\frac{{\mathbf v}^\prime}{c}}\right)\right)+\notag\\
%&\frac{1}{4\pi\epsilon_0}\frac{D^{\prime 3}q}{r^{\prime 2}}\left(1-\frac{v^{\prime 2}}{c^2}\right)\left({\mathbf n}^\prime-{\frac{{\mathbf v}^\prime}{c}}\right),\label{Eformula}\\
%{\mathbf B}=&\frac{1}{4\pi\epsilon_0}\frac{D^{\prime 3}q}{c^3 r^\prime}{\mathbf n}^\prime\times
%\left(-{\mathbf a}^\prime+{\mathbf n}^\prime\times\left({\mathbf a}^\prime\times{\frac{{\mathbf v}^\prime}{c}}\right)\right)+\notag\\
%&\frac{1}{4\pi\epsilon_0}\frac{D^{\prime 3}q}{c r^{\prime 2}}\left(1-\frac{v^{\prime 2}}{c^2}\right)\left({\frac{{\mathbf v}^\prime}{c}}\times{\mathbf n}^\prime\right),\label{Bformula}
%\end{align}
%which satisfy \cite{panofsky1962classical} ${\mathbf B}={\mathbf n}^\prime/c\times {\mathbf E}$. 

The present-time approximations of the Li\'{e}nard-Wiechert fields\cite{griffiths1999introduction,panofsky1962classical,jackson1998classical} up to $1/c^2$ are\cite{o1938electromagnetic, o1965electromagnetic, landau1975classical,jackson1998classical,edwards1976continuing}
\begin{align}
    d{\mathbf E}\approx&\frac{dq}{4\pi\epsilon_0}\left(\frac{\mathbf n}{r^2}+\frac{1}{2}\frac{v^2{\mathbf n}}{r^2 c^2}-\frac{3}{2}\frac{({\mathbf n}\cdot{\mathbf v})^2{\mathbf n}}{r^2c^2}-\frac{\mathbf a}{2rc^2}-\frac{({\mathbf n}\cdot{\mathbf a}){\mathbf n}}{2rc^2}\right),\label{approxforceE}\\
    d{\mathbf B}\approx&\frac{dq}{4\pi\epsilon_0}\frac{{\mathbf v}\times{\mathbf n}}{r^2c^2},\label{approxforceB}
\end{align}
where ${\mathbf n}$, ${\mathbf v}$ and $r$ are present-time quantities, compared to their retarded counterparts ${\mathbf n}^\prime$, ${\mathbf v}^\prime$ and $r^\prime$. The approximate expression of $\mathbf B$ is simple because ${\mathbf B}$ carries $1/c^2$ by itself, and higher-order terms do not survive the approximation.

Note that these approximations are concerning the differences between the corresponding present-time and retarded quantities \emph{of the same charge}, such as those of corresponding arcs in FIG. \ref{fig:byN}(a) and \ref{fig:byN}(b). With this convention, we have pointed out that, in general, ${\mathbf v}\neq{\mathbf v}^\prime$, ${\mathbf r}\neq{\mathbf r}^\prime$ or ${\mathbf n}\neq{\mathbf n}^\prime$, where ${\mathbf v}$, ${\mathbf r}$ and ${\mathbf n}$ are the present-time quantities corresponding to the retarded quantities ${\mathbf v}^\prime$, ${\mathbf r}^\prime$ and ${\mathbf n}^\prime$ \emph{of the same charge}.

Edward, Kenyon, and Lemon \cite{edwards1976continuing} partition terms within parentheses of Eq. (\ref{approxforceE}) into two terms, the first a static term that falls with $1/r^2$ and the second an exact differential. The latter is as follows (Ref. [\onlinecite{edwards1976continuing}, Eq. (7), note that it misses a square symbol])  (Ref. [\onlinecite{mcdonald2019electric},  Eqs. (15) and (16)]),
\begin{align}
    &\frac{dq}{4\pi\epsilon_0}\left(\frac{1}{2}\frac{v^2{\mathbf n}}{r^2c^2}-\frac{3}{2}\frac{({\mathbf n}\cdot{\mathbf v})^2{\mathbf n}}{r^2c^2}-\frac{\mathbf a}{2rc^2}-\frac{({\mathbf n}\cdot{\mathbf a}){\mathbf n}}{2rc^2}\right)\notag\\
    =&\frac{dq}{4\pi\epsilon_0}\left(-\frac{1}{2c^2}\frac{d}{dt}\left(\frac{({\mathbf n}\cdot{\mathbf v}){\mathbf n}}{r}+\frac{\mathbf v}{r}\right)\right).\label{eqn:edward}
\end{align}

The total electric field incurred by the steady loop current is the integral of Eq. (\ref{approxforceE}) over all $dq$'s of the loop current. For a small segment $d{\mathbf l}$ of the loop, with the help of Eq. (\ref{dqdt}), 
Eq. (\ref{eqn:edward}) becomes 
\begin{align}
    \frac{I}{4\pi\epsilon_0}d\left(-\frac{1}{2c^2}\left(\frac{({\mathbf n}\cdot{\mathbf v}){\mathbf n}}{r}+\frac{\mathbf v}{r}\right)\right).\label{eqn:edwardexactd}
\end{align}

Thus, the electric field at position $obs$ is
\begin{align}
{\mathbf E}_{L}\approx
%\approx&\oint \frac{dq}{4\pi\epsilon_0}\left(\frac{\mathbf n}{r^2}+\frac{1}{2}\frac{v^2{\mathbf n}}{r^2 c^2}-\frac{3}{2}\frac{({\mathbf n}\cdot{\mathbf v})^2{\mathbf n}}{r^2c^2}-\frac{\mathbf a}{2rc^2}-\frac{({\mathbf n}\cdot{\mathbf a}){\mathbf n}}{2rc^2}\right)\notag\\
%=\oint \frac{Idt}{4\pi\epsilon_0}\left(\frac{\mathbf n}{r^2}+\frac{1}{2}\frac{v^2{\mathbf n}}{r^2 c^2}-\frac{3}{2}\frac{({\mathbf n}\cdot{\mathbf v})^2{\mathbf n}}{r^2c^2}-\frac{\mathbf a}{2rc^2}-\frac{({\mathbf n}\cdot{\mathbf a}){\mathbf n}}{2rc^2}\right)\notag\\
%=&\frac{I}{4\pi\epsilon_0}\oint \frac{\mathbf n}{r^2} dt + \frac{I}{4\pi\epsilon_0}\oint d\left(-\frac{1}{2c^2}\frac{({\mathbf n}\cdot{\mathbf v}){\mathbf n}}{r}-\frac{1}{2c^2}\frac{\mathbf v}{r}\right)\tag{use Eqs. (\ref{approxforceE}) and (\ref{eqn:edward})}\\
\frac{1}{4\pi\epsilon_0}\oint_L \frac{\mathbf n}{r^2} dq+0=\frac{1}{4\pi\epsilon_0}\oint_L \frac{\mathbf n}{r^2} dq.\label{E_Lapprox}
\end{align}
%where 
%the use of $\oint$ instead of $\int$ for $dt$ emphasizes that the integrands have the same value at the lower and upper limits.  The
%the integration of $dq$ is applied to all $q$ in the loop current. 
%The integration of $dt$ is over the time interval $T$, the total time a charge makes a round trip along the loop. 
Without surprise, this is ordinary Coulomb's law in integration form.

For the magnetic field, with the help of Eq. (\ref{dq}), we have
\begin{align}
{\mathbf B}_{L}\approx&\oint_L \frac{dq}{4\pi\epsilon_0 c^2}\frac{{\mathbf v}\times{\mathbf n}}{r^2}=\oint_L \frac{I}{4\pi\epsilon_0 c^2}\frac{d{\mathbf l}\times{\mathbf n}}{r^2}.\label{B_Lapprox}
\end{align}
This is the ordinary Biot-Savart law in integration form.

\section{Check up our results with the Previous Results}\label{sec:check1}

As depicted in the lower part of FIG. \ref{fig:compare}, as a check-up, with approximations up to $1/c^2$, we expect the first term of Eq. (\ref{dE1}) to equal the first term of Eq. (\ref{approxforceE}) plus an exact differential. That is,
\begin{align}
\frac{{\mathbf n}^\prime}{r^{\prime 2}}D^\prime dq^\prime\approx \frac{\mathbf n}{r^2}dq+(\mbox{an exact differential}),\notag
\end{align}
or, with the help of Eq. (\ref{dq}) and divided by $dq$ or $Idt$,
\begin{align}
\frac{{\mathbf n}^\prime}{r^{\prime 2}}D^\prime \approx \frac{\mathbf n}{r^2}+\frac{d}{dt}(\mbox{a present-time expression}).\label{checkup1}
\end{align}

To check whether this equation holds, we copy out the following approximations from Ritz's work, presented by O'Rahilly \cite{o1965electromagnetic}, with notations used in this paper,
\begin{align}
D^\prime\approx& 1+\frac{{\mathbf n}\cdot{\mathbf v}}{c}+\frac{v^2-r({\mathbf n}\cdot{\mathbf a})}{c^2},\tag{Ref. [\onlinecite{o1965electromagnetic}, p.219, line 17]}\\
r^\prime \approx&  r\left(1+\frac{{\mathbf n}\cdot{\mathbf v}}{c}+\frac{v^2+({\mathbf n}\cdot{\mathbf v})^2-r{\mathbf n}\cdot{\mathbf a}}{2c^2}\right),\tag{Ref. [\onlinecite{o1965electromagnetic}, p.219, line 8]}\\
\frac{1}{r^\prime} \approx& \frac{1}{r}\left(1-\frac{{\mathbf n}\cdot{\mathbf v}}{c}-\frac{v^2-({\mathbf n}\cdot{\mathbf v})^2-r{\mathbf n}\cdot{\mathbf a}}{2c^2}\right).\tag{Ref. [\onlinecite{o1965electromagnetic}, p.219, line 10]}
\end{align}
For people who are familiar with the first-order approximations, it is attempting to write directly the approximation of $1/r^\prime$ as $1/r\left(1-\frac{{\mathbf n}\cdot{\mathbf v}}{c}-\frac{v^2+({\mathbf n}\cdot{\mathbf v})^2-r{\mathbf n}\cdot{\mathbf a}}{2c^2}\right)$, by negating the small-value correction term in the approximation of $r^\prime$. However, the higher-than-first-order approximations are not as simple.

We also write down the following approximations,
\begin{align}
\frac{{\mathbf v}^\prime}{c}=&\frac{1}{c}{\mathbf v}^\prime(t-r^\prime/c) \approx \frac{1}{c}({\mathbf v}-{\mathbf a}\frac{r^\prime}{c})\tag{kinetic equation}\\
\approx& \frac{\mathbf v}{c}-{\mathbf a}\frac{r}{c^2},\tag{use approximate $r^\prime$}\\
{\mathbf 0}-{\mathbf r}^\prime=&{\mathbf 0}-{\mathbf r}^\prime(t-r^\prime/c)\tag{${\mathbf 0}$ is the coordinate of the origin}\\
\approx &{\mathbf 0}-{\mathbf r}+{\mathbf v}\frac{ r^\prime}{c}-\frac{1}{2}{\mathbf a}\frac{r^{\prime 2}}{c^2}\tag{kinetic equation}\\
\approx & {\mathbf 0}-{\mathbf r}+{\mathbf v}\frac{r}{c}\left(1+\frac{{\mathbf n}\cdot{\mathbf v}}{c}\right)-\frac{1}{2}{\mathbf a}\frac{r^2}{c^2}.\tag{use approximate $r^\prime$}
\end{align}

Using these approximations, we get, by throwing out terms of higher order than $1/c^2$, we have
\begin{align}
\frac{1}{r^{\prime 3}}\approx &\frac{1}{r^3}\left(1-3\frac{{\mathbf n}\cdot{\mathbf v}}{c}+3\frac{({\mathbf n}\cdot{\mathbf v})^2}{c^2}-3\frac{v^2-({\mathbf n}\cdot{\mathbf v})^2-r{\mathbf n}\cdot{\mathbf a}}{2c^2}\right),\tag{if $a$ small, $b$ smaller,  $(1+a+b)^3\approx 1+3a+3a^2+3b$}\\
\frac{D^\prime}{r^{\prime 3}}\approx &\frac{1}{r^3}\left(1-3\frac{{\mathbf n}\cdot{\mathbf v}}{c}+3\frac{({\mathbf n}\cdot{\mathbf v})^2}{c^2}-3\frac{v^2-({\mathbf n}\cdot{\mathbf v})^2-r{\mathbf n}\cdot{\mathbf a}}{2c^2}\right.\notag\\
&\left.+\frac{{\mathbf n}\cdot{\mathbf v}}{c}-3\frac{({\mathbf n}\cdot{\mathbf v})^2}{c^2}+\frac{v^2-r({\mathbf n}\cdot{\mathbf a})}{c^2}\right)\tag{use approximate $D^\prime$ and $1/r^{\prime 3}$}\\
= & \frac{1}{r^3}\left(1-2\frac{{\mathbf n}\cdot{\mathbf v}}{c}-\frac{v^2}{2c^2}+\frac{3({\mathbf n}\cdot{\mathbf v})^2}{2c^2}+\frac{r{\mathbf n}\cdot{\mathbf a}}{2c^2}\right),\notag\\
\frac{{\mathbf n}^\prime}{r^{\prime 2}}D^\prime=&\frac{{\mathbf 0}-{\mathbf r}^\prime}{r^{\prime 3}}D^\prime\tag{use ${\mathbf n}^\prime=({\mathbf 0}-{\mathbf r}^\prime)/r^\prime$}\\
\approx & \frac{\mathbf n}{r^2}\left(1-2\frac{{\mathbf n}\cdot{\mathbf v}}{c}-\frac{v^2}{2c^2}+\frac{3({\mathbf n}\cdot{\mathbf v})^2}{2c^2}+\frac{r{\mathbf n}\cdot{\mathbf a}}{2c^2}\right)+\notag\\
&\frac{1}{r^2}\frac{\mathbf v}{c}\left(1+\frac{{\mathbf n}\cdot{\mathbf v}}{c}-2\frac{{\mathbf n}\cdot{\mathbf v}}{c}\right)-\frac{1}{r}\left(\frac{{\mathbf a}}{2 c^2}\right)\tag{use approximate ${D^\prime}/r^{\prime 3}$ and ${\mathbf 0}-{\mathbf r}^\prime$, and ${\mathbf n}=({\mathbf 0}-{\mathbf r})/r$}\\
=&\frac{\mathbf n}{r^2}\left(1-2\frac{{\mathbf n}\cdot{\mathbf v}}{c}-\frac{v^2}{2c^2}+\frac{3({\mathbf n}\cdot{\mathbf v})^2}{2c^2}+\frac{r{\mathbf n}\cdot{\mathbf a}}{2c^2}\right)+\notag\\
&\frac{1}{r^2}\left(\frac{\mathbf v}{c}-\frac{{\mathbf v}({\mathbf n}\cdot{\mathbf v})}{c^2}\right)-\frac{1}{r}\left(\frac{{\mathbf a}}{2 c^2}\right)\tag{combine terms}\\
=&\frac{\mathbf n}{r^2}+\frac{d}{dt}\left(\frac{1}{2c^2}\frac{({\mathbf n}\cdot{\mathbf v}){\mathbf n}}{r}-\frac{1}{2c^2}\frac{\mathbf v}{r}-\frac{1}{c}\frac{\mathbf n}{r}\right), \ \mbox{(see below)}\label{myexact}
\end{align}
where for the last step, we have used the following equations,
\begin{align}
\frac{d}{dt}\left(\frac{({\mathbf n}\cdot{\mathbf v}){\mathbf n}}{r}\right)&=-\frac{v^2{\mathbf n}}{r^2}+\frac{3({\mathbf n}\cdot{\mathbf v})^2{\mathbf n}}{r^2}+\frac{({\mathbf n}\cdot{\mathbf a}){\mathbf n}}{r}-\frac{({\mathbf n}\cdot{\mathbf v}){\mathbf v}}{r^2},\tag{Ref. [\onlinecite{edwards1976continuing}, Eq. (6)]}\\ 
\frac{d}{dt}\left(\frac{\mathbf v}{r}\right)&=\frac{\mathbf a}{r}+\frac{({\mathbf n}\cdot{\mathbf v}){\mathbf v}}{r^2},\tag{Ref. [\onlinecite{edwards1976continuing}, Eq. (5)]}\\
\frac{d}{dt}\left(\frac{\mathbf n}{r}\right)&=\frac{1}{r}\frac{d}{dt}{\mathbf n}+{\mathbf n}\frac{d}{dt}\frac{1}{r}\tag{differentiate by parts}\\
&=\frac{1}{r}\frac{d}{dt}\frac{{\mathbf 0}-{\mathbf r}}{r}+{\mathbf n}\frac{d}{dt}\frac{1}{r}\tag{use ${\mathbf n}=({\mathbf 0}-{\mathbf r})/r$}\\
&=\frac{1}{r}\left(-\frac{\mathbf v}{r}-{\mathbf r}\frac{d}{dt}\frac{1}{r}\right)+{\mathbf n}\frac{d}{dt}\frac{1}{r}\tag{differentiate by parts}\\
&=-\frac{\mathbf v}{r^2}+2{\mathbf n}\frac{d}{dt}\frac{1}{r}\tag{use $-{\mathbf r}/r={\mathbf n}$}\\
&=-\frac{\mathbf v}{r^2}+2{\mathbf n}\frac{d}{dt}({\mathbf r}\cdot{\mathbf r})^{-\frac{1}{2}}\notag\\
&=-\frac{\mathbf v}{r^2}+2{\mathbf n}(-\frac{1}{2r^3})\frac{d}{dt}({\mathbf r}\cdot{\mathbf r})\tag{chain rule}\\
&=-\frac{\mathbf v}{r^2}-2\frac{\mathbf n}{r^3}{\mathbf r}\cdot\frac{d {\mathbf r}}{dt}\tag{differentiate by parts}\\
&=-\frac{\mathbf v}{r^2}+2\frac{({\mathbf n}\cdot{\mathbf v}){\mathbf n}}{r^2}.\tag{use ${\mathbf n}=({\mathbf 0}-{\mathbf r})/r$}
\end{align}

Eq. (\ref{myexact}) is exactly what we have expected at the beginning of this section, shown as Eq. (\ref{checkup1}). Thus, we have reassured ourselves about the correctness of our analysis.

\section{Another check-up}\label{sec:check2}
%To double-check the math, we rewrite Eq. (\ref{E_Lexact}) as
%\begin{align}
%\frac{I}{4\pi\epsilon_0}\oint\frac{{\mathbf n}^\prime}{r^{\prime 2}}D^\prime dt+\frac{I}{4\pi\epsilon_0}\oint \frac{d}{dt}\left(\left({\mathbf n}^\prime-\frac{{\mathbf v}^\prime}{c}\right)\frac{D^\prime}{cr^\prime}\right)dt,\notag
%\end{align}
As depicted in the lower part of FIG. \ref{fig:compare}, as another check-up, if we approximate the second term of Eq. (\ref{dE1}), that is, ${dq}/{(4\pi\epsilon_0)}\frac{d}{dt}\left(\left({\mathbf n}^\prime-\frac{{\mathbf v}^\prime}{c}\right)\frac{D^\prime}{c r^\prime}\right)$, up to $1/c^2$ and combine the result with $dq/(4\pi\epsilon_0)\times$ the second term of Eq. (\ref{myexact}), we expect to recover Eq. (\ref{eqn:edward}) to ``close the loop''. 

Indeed, we have the following approximations,
\begin{align}
{\mathbf n}^\prime=&\frac{{\mathbf 0}-{\mathbf r}^\prime}{r^\prime}\tag{definition of ${\mathbf n}^\prime$}\\
\approx &  {\mathbf n}\left(1-\frac{{\mathbf n}\cdot{\mathbf v}}{c}-\frac{v^2-({\mathbf n}\cdot{\mathbf v})^2-r{\mathbf n}\cdot{\mathbf a}}{2c^2}\right)+\frac{\mathbf v}{c},\tag{use approximate ${\mathbf 0}-{\mathbf r}^\prime$ and $1/r^\prime$}\\
\frac{D^\prime}{c r^\prime}\approx & \frac{1}{c r}.\tag{use approximate $D^\prime$ and $1/r^\prime$; simple for there is $1/c$}
\end{align}
Using these results, we have (simple because there is $1/c$)
\begin{align}
\frac{d}{dt}\left(\left({\mathbf n}^\prime-\frac{{\mathbf v}^\prime}{c}\right)\frac{D^\prime}{c r^\prime}\right) \approx & \frac{d}{dt}\left(\frac{\mathbf n}{c r}\left(1-\frac{{\mathbf n}\cdot{\mathbf v}}{c}\right)\right)\tag{use approximate $D^\prime/(cr^\prime)$, ${\mathbf n}^\prime$ and ${\mathbf v}^\prime/c$}\\
=&\frac{d}{dt}\left(\frac{1}{c}\frac{\mathbf n}{r}-\frac{({\mathbf n}\cdot{\mathbf v}){\mathbf n}}{c^2 r}\right).\label{exactd2}
\end{align}
Comparing this result with the goal of this section, we see that our expectation is met. Again, we have enhanced our confidence that the derivations are correct.

\section{Discussion}
From the moving point charge view, using the fluid model, for an arbitrary steady loop current, we discover an exact differential in the field integrals and show that such a current does not radiate. This is an expected result, but the derivation requires careful work with retardation. We hope our work will shed new light on this century-old result and help students gain a deeper understanding of retardation.

\section{Author Declarations}
\subsection{Conflict of interest}
The authors have no conflict of interest to disclose.

\section{Acknowledgments}

The authors wish to thank the three anonymous reviewers at American Journal of Physics for their very helpful comments and suggestions on an earlier version of this paper.

\section{Appendix: Code}\label{sec:code}

The following is the GNU octave code (Matlab compatible) used in plotting FIG. \ref{fig:byN}.
{\small
\begin{verbatim}
#Octave code  to calculate retarded times for 
#labeled charges in a circlar same-speed steady
#current. 
#GNU Octave is downloadable from its website. 
#There are websites that run octave code on-line. 
#The code also runs in Matlab.
a = 1; #a is radius in length unit of lightsecond
cc = 1; #cc is light speed in lightsecond/second 
v = 0.99*cc; #Charge speed; 0.99cc to exaggerate
T = 2*pi*a/v; w = 2*pi/T; #Period and angular speed
N=16; #number of labeled charges

indice=transpose(linspace(0, N-1, N));
phi = (2*pi/N)*indice;
#Phases of the N charges evenly spreaded 
#in [0, 2*pi]
tr = zeros(N, 1); #for storing retarded times
for i = 1:length(phi)  #for loop to calculate tr
  t1 = 0; #The initial value for the iteration
  for j = 1:20 #20 iterations enough to converge
    t1 = 0 - a*sqrt((a/2-cos(w*t1+phi(i)))^2+ \
        (0-sin(w*t1+phi(i)))^2)/cc;
    #It is the eqn that defines the retarded time:
    #tr = t-|r_O-r_'(tr)|/c
    #The observer is at position (a/2,0); and t=0
    #t1   #uncomment to print t1 during iteration
  end
  tr(i) = t1; #store the calculated retarded time
end

x = a*cos(phi); y = a*sin(phi); #Present positions
lx =1.1*x; ly=1.1*y; #For printing labels 
xr =a*cos(w*tr+phi); yr = a*sin(w*tr+phi); 
  #retarded positions
lxr =1.1*xr; lyr = 1.1*yr; #For printing labels
data=[indice,-1.2+x,y,-1.2+lx,ly, \
  1.2+xr,yr,1.2+lxr,lyr] #Offset with -1.2, 1.2
save data.txt data 
\end{verbatim}}

The following is the Gnuplot code used in generating the graphs used in FIG. \ref{fig:byN}.
{\small
\begin{verbatim}
#Gnuplot code to plot data in the file data.txt.
#There are on-line websites that run gnuplot
#code on-line.

#Output to xfig format to add annotatations,
set terminal fig; set output "graph.fig"

#Uncomment this line to write to png format,
#set terminal pngcairo; set output "graph.png"

unset key  #Remove default legends
set title "" font "Times, 10" #No title
set xrange[-2.4:2.4]
set yrange[-1.2:1.2]
unset border  #No border
unset xtics 
unset ytics
set size ratio 0.5 #height:width 

#Draw two circles,
set object 1 circle front at -1.2,0 size 1 \
  fillcolor rgb "black" lw 1
set object 2 circle front at 1.2,0 size 1 \
  fillcolor rgb "black" lw 1

plot "data.txt" \  #Plot columns in data.txt
     using 2:3 with points pointtype 7 \
     lc rgb "black",\
  "" using 4:5:1 with labels font "Times, 12",\
  "" using 6:7 with points pointtype 7 \
     lc rgb "black",\
  "" using 8:9:1 with labels font "Times, 12"
\end{verbatim}}

\nocite{*}
\bibliography{reference_steadyloop}

\end{document}

%% file: compare.pdf_t
\begin{picture}(0,0)%
\includegraphics{compare.pdf}%
\end{picture}%
\setlength{\unitlength}{3947sp}%
\begingroup\makeatletter\ifx\SetFigFont\undefined%
\gdef\SetFigFont#1#2#3#4#5{%
  \reset@font\fontsize{#1}{#2pt}%
  \fontfamily{#3}\fontseries{#4}\fontshape{#5}%
  \selectfont}%
\fi\endgroup%
\begin{picture}(5554,8406)(879,-9280)
\put(976,-2761){\makebox(0,0)[lb]{\smash{{\SetFigFont{12}{14.4}{\rmdefault}{\mddefault}{\updefault}{\color[rgb]{0,0,0}Present-time E field\cite{o1938electromagnetic,o1965electromagnetic,landau1975classical,jackson1998classical,edwards1976continuing}}%
}}}}
\put(976,-2911){\makebox(0,0)[lb]{\smash{{\SetFigFont{12}{14.4}{\rmdefault}{\mddefault}{\updefault}{\color[rgb]{0,0,0}Eq. (\ref{approxforceE})}%
}}}}
\put(976,-1936){\makebox(0,0)[lb]{\smash{{\SetFigFont{12}{14.4}{\rmdefault}{\mddefault}{\updefault}{\color[rgb]{0,0,0}Present-time potentials}%
}}}}
\put(1651,-2236){\makebox(0,0)[lb]{\smash{{\SetFigFont{12}{14.4}{\rmdefault}{\mddefault}{\updefault}{\color[rgb]{0,0,0}Differentiate}%
}}}}
\put(1651,-1411){\makebox(0,0)[lb]{\smash{{\SetFigFont{12}{14.4}{\rmdefault}{\mddefault}{\updefault}{\color[rgb]{0,0,0}Approximate up to $1/c^2$}%
}}}}
\put(1651,-1636){\makebox(0,0)[lb]{\smash{{\SetFigFont{12}{14.4}{\rmdefault}{\mddefault}{\updefault}{\color[rgb]{0,0,0}(O'Rahilly\cite{o1938electromagnetic,o1965electromagnetic})}%
}}}}
\put(1651,-3211){\makebox(0,0)[lb]{\smash{{\SetFigFont{12}{14.4}{\rmdefault}{\mddefault}{\updefault}{\color[rgb]{0,0,0}Partition\cite{edwards1976continuing}}%
}}}}
\put(4126,-2761){\makebox(0,0)[lb]{\smash{{\SetFigFont{12}{14.4}{\rmdefault}{\mddefault}{\updefault}{\color[rgb]{0,0,0}Li\'{e}nard-Wiechert fields\cite{griffiths1999introduction,panofsky1962classical,jackson1998classical}}%
}}}}
\put(5026,-2236){\makebox(0,0)[lb]{\smash{{\SetFigFont{12}{14.4}{\rmdefault}{\mddefault}{\updefault}{\color[rgb]{0,0,0}Differentiate\cite{griffiths1999introduction,panofsky1962classical,jackson1998classical} }%
}}}}
\put(5026,-3211){\makebox(0,0)[lb]{\smash{{\SetFigFont{12}{14.4}{\rmdefault}{\mddefault}{\updefault}{\color[rgb]{0,0,0}Partition}%
}}}}
\put(5026,-3361){\makebox(0,0)[lb]{\smash{{\SetFigFont{12}{14.4}{\rmdefault}{\mddefault}{\updefault}{\color[rgb]{0,0,0}(Sect. \ref{sec:Fields})}%
}}}}
\put(1126,-5236){\rotatebox{90.0}{\makebox(0,0)[lb]{\smash{{\SetFigFont{12}{14.4}{\rmdefault}{\mddefault}{\updefault}{\color[rgb]{0,0,0}A present-time}%
}}}}}
\put(1276,-5236){\rotatebox{90.0}{\makebox(0,0)[lb]{\smash{{\SetFigFont{12}{14.4}{\rmdefault}{\mddefault}{\updefault}{\color[rgb]{0,0,0}exact differential\cite{edwards1976continuing}}%
}}}}}
\put(1426,-5236){\rotatebox{90.0}{\makebox(0,0)[lb]{\smash{{\SetFigFont{12}{14.4}{\rmdefault}{\mddefault}{\updefault}{\color[rgb]{0,0,0}for segment $d{\mathbf l}$;}%
}}}}}
\put(1576,-5236){\rotatebox{90.0}{\makebox(0,0)[lb]{\smash{{\SetFigFont{12}{14.4}{\rmdefault}{\mddefault}{\updefault}{\color[rgb]{0,0,0}Eq. (\ref{eqn:edward})}%
}}}}}
\put(1951,-5236){\rotatebox{90.0}{\makebox(0,0)[lb]{\smash{{\SetFigFont{12}{14.4}{\rmdefault}{\mddefault}{\updefault}{\color[rgb]{0,0,0}Present-time}%
}}}}}
\put(2101,-5236){\rotatebox{90.0}{\makebox(0,0)[lb]{\smash{{\SetFigFont{12}{14.4}{\rmdefault}{\mddefault}{\updefault}{\color[rgb]{0,0,0}Coulomb's law}%
}}}}}
\put(2251,-5236){\rotatebox{90.0}{\makebox(0,0)[lb]{\smash{{\SetFigFont{12}{14.4}{\rmdefault}{\mddefault}{\updefault}{\color[rgb]{0,0,0}for segment $d{\mathbf l}$;}%
}}}}}
\put(2401,-5236){\rotatebox{90.0}{\makebox(0,0)[lb]{\smash{{\SetFigFont{12}{14.4}{\rmdefault}{\mddefault}{\updefault}{\color[rgb]{0,0,0}1st term of Eq. (\ref{approxforceE})}%
}}}}}
\put(2776,-5011){\rotatebox{90.0}{\makebox(0,0)[lb]{\smash{{\SetFigFont{12}{14.4}{\rmdefault}{\mddefault}{\updefault}{\color[rgb]{0,0,0}Equal E integral over}%
}}}}}
\put(2926,-5011){\rotatebox{90.0}{\makebox(0,0)[lb]{\smash{{\SetFigFont{12}{14.4}{\rmdefault}{\mddefault}{\updefault}{\color[rgb]{0,0,0}a steady loop current}%
}}}}}
\put(3376,-5236){\rotatebox{90.0}{\makebox(0,0)[lb]{\smash{{\SetFigFont{12}{14.4}{\rmdefault}{\mddefault}{\updefault}{\color[rgb]{0,0,0}\emph{Present-time}}%
}}}}}
\put(3526,-5236){\rotatebox{90.0}{\makebox(0,0)[lb]{\smash{{\SetFigFont{12}{14.4}{\rmdefault}{\mddefault}{\updefault}{\color[rgb]{0,0,0}Coulomb's law and}%
}}}}}
\put(3676,-5236){\rotatebox{90.0}{\makebox(0,0)[lb]{\smash{{\SetFigFont{12}{14.4}{\rmdefault}{\mddefault}{\updefault}{\color[rgb]{0,0,0}Biot-Savart law}%
}}}}}
\put(3976,-5236){\rotatebox{90.0}{\makebox(0,0)[lb]{\smash{{\SetFigFont{12}{14.4}{\rmdefault}{\mddefault}{\updefault}{\color[rgb]{0,0,0}Eqs. (\ref{retardedCoulomb}), (\ref{retardedBfield})}%
}}}}}
\put(4426,-5011){\rotatebox{90.0}{\makebox(0,0)[lb]{\smash{{\SetFigFont{12}{14.4}{\rmdefault}{\mddefault}{\updefault}{\color[rgb]{0,0,0}(Sect. \ref{sec:Fields})}%
}}}}}
\put(5026,-5236){\rotatebox{90.0}{\makebox(0,0)[lb]{\smash{{\SetFigFont{12}{14.4}{\rmdefault}{\mddefault}{\updefault}{\color[rgb]{0,0,0}Biot-Savart law}%
}}}}}
\put(5176,-5236){\rotatebox{90.0}{\makebox(0,0)[lb]{\smash{{\SetFigFont{12}{14.4}{\rmdefault}{\mddefault}{\updefault}{\color[rgb]{0,0,0}for segment $d{\mathbf l}^\prime$; 1st}%
}}}}}
\put(5326,-5236){\rotatebox{90.0}{\makebox(0,0)[lb]{\smash{{\SetFigFont{12}{14.4}{\rmdefault}{\mddefault}{\updefault}{\color[rgb]{0,0,0}terms of (\ref{dE1}), (\ref{dB1})}%
}}}}}
\put(5701,-5236){\rotatebox{90.0}{\makebox(0,0)[lb]{\smash{{\SetFigFont{12}{14.4}{\rmdefault}{\mddefault}{\updefault}{\color[rgb]{0,0,0}Retarded exact}%
}}}}}
\put(5851,-5236){\rotatebox{90.0}{\makebox(0,0)[lb]{\smash{{\SetFigFont{12}{14.4}{\rmdefault}{\mddefault}{\updefault}{\color[rgb]{0,0,0}differentials for}%
}}}}}
\put(6001,-5236){\rotatebox{90.0}{\makebox(0,0)[lb]{\smash{{\SetFigFont{12}{14.4}{\rmdefault}{\mddefault}{\updefault}{\color[rgb]{0,0,0}segment $d{\mathbf l}^\prime$; 2nd}%
}}}}}
\put(6151,-5236){\rotatebox{90.0}{\makebox(0,0)[lb]{\smash{{\SetFigFont{12}{14.4}{\rmdefault}{\mddefault}{\updefault}{\color[rgb]{0,0,0}terms of (\ref{dE1}), (\ref{dB1})}%
}}}}}
\put(3076,-5011){\rotatebox{90.0}{\makebox(0,0)[lb]{\smash{{\SetFigFont{12}{14.4}{\rmdefault}{\mddefault}{\updefault}{\color[rgb]{0,0,0}(Sect. \ref{sec:Fields})}%
}}}}}
\put(3226,-2686){\makebox(0,0)[lb]{\smash{{\SetFigFont{12}{14.4}{\rmdefault}{\mddefault}{\updefault}{\color[rgb]{0,0,0}are done\cite{mcdonald2019electric}}%
}}}}
\put(3226,-1936){\makebox(0,0)[lb]{\smash{{\SetFigFont{12}{14.4}{\rmdefault}{\mddefault}{\updefault}{\color[rgb]{0,0,0}Approximate E field}%
}}}}
\put(3226,-2461){\makebox(0,0)[lb]{\smash{{\SetFigFont{12}{14.4}{\rmdefault}{\mddefault}{\updefault}{\color[rgb]{0,0,0}Acceleration terms}%
}}}}
\put(3226,-2161){\makebox(0,0)[lb]{\smash{{\SetFigFont{12}{14.4}{\rmdefault}{\mddefault}{\updefault}{\color[rgb]{0,0,0}up to $1/c^2$}%
}}}}
\put(3226,-2311){\makebox(0,0)[lb]{\smash{{\SetFigFont{12}{14.4}{\rmdefault}{\mddefault}{\updefault}{\color[rgb]{0,0,0}Can be an excise}%
}}}}
\put(5326,-5911){\makebox(0,0)[lb]{\smash{{\SetFigFont{12}{14.4}{\rmdefault}{\mddefault}{\updefault}{\color[rgb]{0,0,0}(Sect. \ref{sec:check2})}%
}}}}
\put(2251,-1111){\makebox(0,0)[lb]{\smash{{\SetFigFont{12}{14.4}{\rmdefault}{\mddefault}{\updefault}{\color[rgb]{0,0,0}Li\'{e}nard-Wiechert potentials\cite{griffiths1999introduction,panofsky1962classical,jackson1998classical} }%
}}}}
\put(1651,-2461){\makebox(0,0)[lb]{\smash{{\SetFigFont{12}{14.4}{\rmdefault}{\mddefault}{\updefault}{\color[rgb]{0,0,0}(O'Rahilly\cite{o1938electromagnetic,o1965electromagnetic})}%
}}}}
\put(4876,-5236){\rotatebox{90.0}{\makebox(0,0)[lb]{\smash{{\SetFigFont{12}{14.4}{\rmdefault}{\mddefault}{\updefault}{\color[rgb]{0,0,0}Coulomb's law and}%
}}}}}
\put(3226,-2911){\makebox(0,0)[lb]{\smash{{\SetFigFont{12}{14.4}{\rmdefault}{\mddefault}{\updefault}{\color[rgb]{0,0,0}for the $x$}%
}}}}
\put(3226,-3061){\makebox(0,0)[lb]{\smash{{\SetFigFont{12}{14.4}{\rmdefault}{\mddefault}{\updefault}{\color[rgb]{0,0,0}axis}%
}}}}
\put(4276,-5011){\rotatebox{90.0}{\makebox(0,0)[lb]{\smash{{\SetFigFont{12}{14.4}{\rmdefault}{\mddefault}{\updefault}{\color[rgb]{0,0,0}Equal}%
}}}}}
\put(4726,-5236){\rotatebox{90.0}{\makebox(0,0)[lb]{\smash{{\SetFigFont{12}{14.4}{\rmdefault}{\mddefault}{\updefault}{\color[rgb]{0,0,0}Retarded}%
}}}}}
\put(4126,-5911){\makebox(0,0)[lb]{\smash{{\SetFigFont{12}{14.4}{\rmdefault}{\mddefault}{\updefault}{\color[rgb]{0,0,0}and partition}%
}}}}
\put(5326,-5461){\makebox(0,0)[lb]{\smash{{\SetFigFont{12}{14.4}{\rmdefault}{\mddefault}{\updefault}{\color[rgb]{0,0,0}Approximate}%
}}}}
\put(5326,-5611){\makebox(0,0)[lb]{\smash{{\SetFigFont{12}{14.4}{\rmdefault}{\mddefault}{\updefault}{\color[rgb]{0,0,0}E field}%
}}}}
\put(5326,-5761){\makebox(0,0)[lb]{\smash{{\SetFigFont{12}{14.4}{\rmdefault}{\mddefault}{\updefault}{\color[rgb]{0,0,0}up to $1/c^2$}%
}}}}
\put(4126,-5461){\makebox(0,0)[lb]{\smash{{\SetFigFont{12}{14.4}{\rmdefault}{\mddefault}{\updefault}{\color[rgb]{0,0,0}Approximate}%
}}}}
\put(4126,-5761){\makebox(0,0)[lb]{\smash{{\SetFigFont{12}{14.4}{\rmdefault}{\mddefault}{\updefault}{\color[rgb]{0,0,0}up to $1/c^2$}%
}}}}
\put(4126,-5611){\makebox(0,0)[lb]{\smash{{\SetFigFont{12}{14.4}{\rmdefault}{\mddefault}{\updefault}{\color[rgb]{0,0,0}E field}%
}}}}
\put(1276,-9211){\makebox(0,0)[lb]{\smash{{\SetFigFont{12}{14.4}{\rmdefault}{\mddefault}{\updefault}{\color[rgb]{0,0,0}: New results in this paper}%
}}}}
\put(1276,-8986){\makebox(0,0)[lb]{\smash{{\SetFigFont{12}{14.4}{\rmdefault}{\mddefault}{\updefault}{\color[rgb]{0,0,0}: Approximate quantities}%
}}}}
\put(4876,-8236){\rotatebox{90.0}{\makebox(0,0)[lb]{\smash{{\SetFigFont{12}{14.4}{\rmdefault}{\mddefault}{\updefault}{\color[rgb]{0,0,0}A present-time}%
}}}}}
\put(5026,-8236){\rotatebox{90.0}{\makebox(0,0)[lb]{\smash{{\SetFigFont{12}{14.4}{\rmdefault}{\mddefault}{\updefault}{\color[rgb]{0,0,0}exact differential for}%
}}}}}
\put(5701,-8236){\rotatebox{90.0}{\makebox(0,0)[lb]{\smash{{\SetFigFont{12}{14.4}{\rmdefault}{\mddefault}{\updefault}{\color[rgb]{0,0,0}A present-time}%
}}}}}
\put(5851,-8236){\rotatebox{90.0}{\makebox(0,0)[lb]{\smash{{\SetFigFont{12}{14.4}{\rmdefault}{\mddefault}{\updefault}{\color[rgb]{0,0,0}exact differential}%
}}}}}
\put(4201,-8236){\rotatebox{90.0}{\makebox(0,0)[lb]{\smash{{\SetFigFont{12}{14.4}{\rmdefault}{\mddefault}{\updefault}{\color[rgb]{0,0,0}Coulomb's law for}%
}}}}}
\put(4051,-8236){\rotatebox{90.0}{\makebox(0,0)[lb]{\smash{{\SetFigFont{12}{14.4}{\rmdefault}{\mddefault}{\updefault}{\color[rgb]{0,0,0}Present-time}%
}}}}}
\put(5101,-8536){\makebox(0,0)[lb]{\smash{{\SetFigFont{12}{14.4}{\rmdefault}{\mddefault}{\updefault}{\color[rgb]{0,0,0}Sum up}%
}}}}
\put(1276,-8536){\makebox(0,0)[lb]{\smash{{\SetFigFont{12}{14.4}{\rmdefault}{\mddefault}{\updefault}{\color[rgb]{0,0,0}(Sect. \ref{sec:check2})}%
}}}}
\put(4351,-8236){\rotatebox{90.0}{\makebox(0,0)[lb]{\smash{{\SetFigFont{12}{14.4}{\rmdefault}{\mddefault}{\updefault}{\color[rgb]{0,0,0}segment $d{\mathbf l}$; $dq/(4\pi\epsilon_0)\times$}%
}}}}}
\put(4501,-8236){\rotatebox{90.0}{\makebox(0,0)[lb]{\smash{{\SetFigFont{12}{14.4}{\rmdefault}{\mddefault}{\updefault}{\color[rgb]{0,0,0}1st term of Eq. (\ref{myexact})}%
}}}}}
\put(5176,-8236){\rotatebox{90.0}{\makebox(0,0)[lb]{\smash{{\SetFigFont{12}{14.4}{\rmdefault}{\mddefault}{\updefault}{\color[rgb]{0,0,0}segment $d{\mathbf l}$; $dq/(4\pi\epsilon_0)\times$}%
}}}}}
\put(5326,-8236){\rotatebox{90.0}{\makebox(0,0)[lb]{\smash{{\SetFigFont{12}{14.4}{\rmdefault}{\mddefault}{\updefault}{\color[rgb]{0,0,0}2nd term of Eq. (\ref{myexact})}%
}}}}}
\put(6001,-8236){\rotatebox{90.0}{\makebox(0,0)[lb]{\smash{{\SetFigFont{12}{14.4}{\rmdefault}{\mddefault}{\updefault}{\color[rgb]{0,0,0}for segment $d{\mathbf l}$;}%
}}}}}
\put(6151,-8236){\rotatebox{90.0}{\makebox(0,0)[lb]{\smash{{\SetFigFont{12}{14.4}{\rmdefault}{\mddefault}{\updefault}{\color[rgb]{0,0,0}$dq/(4\pi\epsilon_0)\times$ Eq. (\ref{exactd2})}%
}}}}}
\put(4201,-6136){\makebox(0,0)[lb]{\smash{{\SetFigFont{12}{14.4}{\rmdefault}{\mddefault}{\updefault}{\color[rgb]{0,0,0}(Sect. \ref{sec:check1})}%
}}}}
\put(1951,-7036){\makebox(0,0)[lb]{\smash{{\SetFigFont{12}{14.4}{\rmdefault}{\mddefault}{\updefault}{\color[rgb]{0,0,0}A check-up}%
}}}}
\put(1951,-7186){\makebox(0,0)[lb]{\smash{{\SetFigFont{12}{14.4}{\rmdefault}{\mddefault}{\updefault}{\color[rgb]{0,0,0}shall be equal}%
}}}}
\put(1951,-7336){\makebox(0,0)[lb]{\smash{{\SetFigFont{12}{14.4}{\rmdefault}{\mddefault}{\updefault}{\color[rgb]{0,0,0}(Sect. \ref{sec:check1})}%
}}}}
\put(1276,-8236){\makebox(0,0)[lb]{\smash{{\SetFigFont{12}{14.4}{\rmdefault}{\mddefault}{\updefault}{\color[rgb]{0,0,0}Another check-up}%
}}}}
\put(1276,-8386){\makebox(0,0)[lb]{\smash{{\SetFigFont{12}{14.4}{\rmdefault}{\mddefault}{\updefault}{\color[rgb]{0,0,0}shall be equal}%
}}}}
\put(3826,-5236){\rotatebox{90.0}{\makebox(0,0)[lb]{\smash{{\SetFigFont{12}{14.4}{\rmdefault}{\mddefault}{\updefault}{\color[rgb]{0,0,0}for segment $d{\mathbf l}^\prime$; }%
}}}}}
\end{picture}%

%% file: graph_edit.pdf_t
\begin{picture}(0,0)%
\includegraphics{graph_edit.pdf}%
\end{picture}%
%
%  Produced by gnuplot version 5.2 
%
\setlength{\unitlength}{3947sp}%
\begingroup\makeatletter\ifx\SetFigFont\undefined%
\gdef\SetFigFont#1#2#3#4#5{%
  \reset@font\fontsize{#1}{#2pt}%
  \fontfamily{#3}\fontseries{#4}\fontshape{#5}%
  \selectfont}%
\fi\endgroup%
\begin{picture}(5621,3155)(1442,-3507)
\put(3676,-3436){\makebox(0,0)[lb]{\smash{{\SetFigFont{12}{14.4}{\rmdefault}{\mddefault}{\updefault}{\color[rgb]{0,0,0}$I$}%
}}}}
\put(6526,-3436){\makebox(0,0)[lb]{\smash{{\SetFigFont{12}{14.4}{\rmdefault}{\mddefault}{\updefault}{\color[rgb]{0,0,0}$I$}%
}}}}
\put(1501,-511){\makebox(0,0)[lb]{\smash{{\SetFigFont{12}{14.4}{\rmdefault}{\mddefault}{\updefault}{\color[rgb]{0,0,0}(a) Present-time positions of $N$}%
}}}}
\put(4351,-511){\makebox(0,0)[lb]{\smash{{\SetFigFont{12}{14.4}{\rmdefault}{\mddefault}{\updefault}{\color[rgb]{0,0,0}(b) Retarded positions of $N$}%
}}}}
\put(4651,-661){\makebox(0,0)[lb]{\smash{{\SetFigFont{12}{14.4}{\rmdefault}{\mddefault}{\updefault}{\color[rgb]{0,0,0}labeled point charges, $N=16$}%
}}}}
\put(6076,-2386){\makebox(0,0)[lb]{\smash{{\SetFigFont{12}{14.4}{\rmdefault}{\mddefault}{\updefault}{\color[rgb]{0,0,0}$obs$}%
}}}}
\put(1801,-661){\makebox(0,0)[lb]{\smash{{\SetFigFont{12}{14.4}{\rmdefault}{\mddefault}{\updefault}{\color[rgb]{0,0,0}labeled point charges, $N=16$}%
}}}}
\end{picture}%

%% file: closedcircuitnewcounterclockwise.pdf_t
\begin{picture}(0,0)%
\includegraphics{closedcircuitnewcounterclockwise.pdf}%
\end{picture}%
\setlength{\unitlength}{3947sp}%
\begingroup\makeatletter\ifx\SetFigFont\undefined%
\gdef\SetFigFont#1#2#3#4#5{%
  \reset@font\fontsize{#1}{#2pt}%
  \fontfamily{#3}\fontseries{#4}\fontshape{#5}%
  \selectfont}%
\fi\endgroup%
\begin{picture}(4530,2544)(1336,-2221)
\put(5851,-736){\makebox(0,0)[lb]{\smash{{\SetFigFont{12}{14.4}{\rmdefault}{\mddefault}{\updefault}{\color[rgb]{0,0,0}${\mathbf r}_{k-1}^\prime$}%
}}}}
\put(5776,-436){\makebox(0,0)[lb]{\smash{{\SetFigFont{12}{14.4}{\rmdefault}{\mddefault}{\updefault}{\color[rgb]{0,0,0}$\delta {\mathbf l}_k^\prime\Rightarrow d{\mathbf l}^\prime$}%
}}}}
\put(5551,-211){\makebox(0,0)[lb]{\smash{{\SetFigFont{12}{14.4}{\rmdefault}{\mddefault}{\updefault}{\color[rgb]{0,0,0}${\mathbf r}_k^\prime\Rightarrow{\mathbf r}^\prime$}%
}}}}
\put(5326,-2086){\makebox(0,0)[lb]{\smash{{\SetFigFont{12}{14.4}{\rmdefault}{\mddefault}{\updefault}{\color[rgb]{0,0,0}$I$}%
}}}}
\put(5251,-1111){\makebox(0,0)[lb]{\smash{{\SetFigFont{12}{14.4}{\rmdefault}{\mddefault}{\updefault}{\color[rgb]{0,0,0}$obs$}%
}}}}
\put(4726,-1186){\makebox(0,0)[lb]{\smash{{\SetFigFont{12}{14.4}{\rmdefault}{\mddefault}{\updefault}{\color[rgb]{0,0,0}${\mathbf r}_{obs}$}%
}}}}
\put(1501,-2011){\makebox(0,0)[lb]{\smash{{\SetFigFont{12}{14.4}{\rmdefault}{\mddefault}{\updefault}{\color[rgb]{0,0,0}$I$}%
}}}}
\put(2776,-2161){\makebox(0,0)[lb]{\smash{{\SetFigFont{12}{14.4}{\rmdefault}{\mddefault}{\updefault}{\color[rgb]{0,0,0}$I$}%
}}}}
\put(1801,-1336){\makebox(0,0)[lb]{\smash{{\SetFigFont{12}{14.4}{\rmdefault}{\mddefault}{\updefault}{\color[rgb]{0,0,0}$O$}%
}}}}
\put(2926,-1561){\makebox(0,0)[lb]{\smash{{\SetFigFont{12}{14.4}{\rmdefault}{\mddefault}{\updefault}{\color[rgb]{0,0,0}${\mathbf r}_{k-1}$}%
}}}}
\put(3151,-61){\makebox(0,0)[lb]{\smash{{\SetFigFont{12}{14.4}{\rmdefault}{\mddefault}{\updefault}{\color[rgb]{0,0,0}The two segments}%
}}}}
\put(3151,-211){\makebox(0,0)[lb]{\smash{{\SetFigFont{12}{14.4}{\rmdefault}{\mddefault}{\updefault}{\color[rgb]{0,0,0}contain the}%
}}}}
\put(3151,-361){\makebox(0,0)[lb]{\smash{{\SetFigFont{12}{14.4}{\rmdefault}{\mddefault}{\updefault}{\color[rgb]{0,0,0}same charge}%
}}}}
\put(3151,-1336){\makebox(0,0)[lb]{\smash{{\SetFigFont{12}{14.4}{\rmdefault}{\mddefault}{\updefault}{\color[rgb]{0,0,0}$\delta {\mathbf l}_k\Rightarrow d{\mathbf l}$}%
}}}}
\put(3301,-1111){\makebox(0,0)[lb]{\smash{{\SetFigFont{12}{14.4}{\rmdefault}{\mddefault}{\updefault}{\color[rgb]{0,0,0}${\mathbf r}_k\Rightarrow{\mathbf r}$}%
}}}}
\put(4726,-211){\makebox(0,0)[lb]{\smash{{\SetFigFont{12}{14.4}{\rmdefault}{\mddefault}{\updefault}{\color[rgb]{0,0,0}$I$}%
}}}}
\put(2176,-286){\makebox(0,0)[lb]{\smash{{\SetFigFont{12}{14.4}{\rmdefault}{\mddefault}{\updefault}{\color[rgb]{0,0,0}$I$}%
}}}}
\put(4126,-2011){\makebox(0,0)[lb]{\smash{{\SetFigFont{12}{14.4}{\rmdefault}{\mddefault}{\updefault}{\color[rgb]{0,0,0}$I$}%
}}}}
\put(4351,-1261){\makebox(0,0)[lb]{\smash{{\SetFigFont{12}{14.4}{\rmdefault}{\mddefault}{\updefault}{\color[rgb]{0,0,0}$O$}%
}}}}
\put(1351,164){\makebox(0,0)[lb]{\smash{{\SetFigFont{12}{14.4}{\rmdefault}{\mddefault}{\updefault}{\color[rgb]{0,0,0}(a)Present-time segment $d{\mathbf l}_k$}%
}}}}
\put(3901,164){\makebox(0,0)[lb]{\smash{{\SetFigFont{12}{14.4}{\rmdefault}{\mddefault}{\updefault}{\color[rgb]{0,0,0}(b)Retarded segment $d{\mathbf l}_k^\prime$}%
}}}}
\end{picture}%